\documentclass[preprint,12pt]{elsarticle}
\usepackage{amssymb}
\usepackage{graphicx}
\usepackage{amsmath} 
\usepackage{amssymb}
\usepackage{amsfonts}
\usepackage{mathrsfs}
\usepackage{units}
\usepackage{multirow}
\usepackage{dcolumn}
\usepackage{bm}
\usepackage{rotating} 
\newcommand{\bs}[1]{\boldsymbol{#1}}
\journal{Physics Letters B}

\begin{document}
\begin{frontmatter}

\title{Coupled-channel continuum eigenchannel basis}

\author{R. M. Id Betan}
\ead{idbetan@ifir-conicet.gov.ar}
\address{Physics Institute of Rosario (CONICET),\\
         Bvrd. 27 de Febrero 210 bis, S2000FPA Rosario, Argentina}
\address{Department of Physics and Chemistry (FCEIA-UNR),\\
         Av. Pellegrini 250, S2000BTP Rosario, Argentina}
\date{\today}
\begin{abstract} 
The goal of this paper is  to calculate bound, resonant and scattering states in the coupled-channel formalism without relying on the boundary conditions at large distances. The coupled-channel solution is expanded in eigenchannel bases i.e. in eigenfunctions of diagonal Hamiltonians. Each eigenchannel basis may include  discrete  and discretized continuum (real or complex energy) single particle states. The coupled-channel solutions are computed through diagonalization in these bases. The method is applied to a few two-channels problems.
The exact bound spectrum of the Poeschl-Teller potential is well described by using a basis of real energy continuum states. For deuteron described by Reid potential, the experimental energy and the $S$ and $D$  contents of the wave function  are reproduced in the asymptotic limit of the cutoff energy. For the Noro-Taylor potential resonant state energy is well reproduced by using the complex energy Berggren basis. It is found that the expansion of the coupled-channel wave function in these eigenchannel bases require less computational efforts than the use of any other basis. The solutions are stable and converge as the cutoff energy  increases.
\end{abstract}
\end{frontmatter}

\section{Introduction} \label{sec.introduction}
Considerable amount of efforts are devoted all around the world for studying the properties of unstable nuclei \cite{2013Tanihata}. Because of this, new theoretical approaches, which takes into account the continuum explicitly, is called for revealing their properties. The coupled-channel method is a very powerful formalism for studying the structure of both strongly-bound nuclei \cite{1996Frobrich,1988Thompson} and loosely-bound nuclei \cite{2012Upadhyay} too. Here we propose a way to calculate the coupled-channel solutions in which all bound and continuum (resonant and non-resonant continuum) states are treated on the equal footing.

Complex eigenenergies, i.e. Gamow \cite{1928Gamow} or Siegert \cite{1939Siegert} states were calculated using the Green's function approach in momentum 
space in Refs.  \cite{1983Landau,1989He} for coupled channel problems. Gamow states for realistic deformed potentials were calculated first in Ref. \cite{1997Ferreira} by solving the logarithmic derivative of the coupled equations with outgoing boundary condition. In Refs. \cite{1999Rykaczewski} and \cite{2000Kruppa} the coupled-channel Schr\"odinger equation with outgoing wave boundary condition were used to study the proton decay states in a rare-earth nucleus.

The complex scaling method has been successfully combined with the coupled-equation formalism to calculate resonances \cite{1998Rakityanasky,1999Masui,1999Kato,2013Dote}. The extension of the Gamow Shell Model \cite{2002IdBetan,2002Michel} to reaction problems in the framework of coupled-channel formalism was recently implemented in Ref. \cite{2012Jaganathen}, where the low-lying states of $^7$Li were calculated. The result of the direct integration of coupled equations was compared with that of the Berggren \cite{1968Berggren} expansion for the calculation of bound states of dipolar molecules in Ref. \cite{2013Fossez}. A full complex energy representation, was used  in Ref. \cite{2008IdBetan} for the calculation of the Isobaric Analog State by coupled Lane equations. The present paper extends the use of the continuum bases to the inelastic processes in coupled-equation systems and to the calculation of scattering states.

The method presented in this paper allows the calculation of bound, resonances and scattering states in coupled systems on the same footing. All these states may be found by a single diagonalization simultaneously. Each channel wave function is expanded in an optimized basis set defined by the eigensolution of the corresponding uncoupled Schr\"odinger equation, that is the meaning of eigenchannel bases. Since this method prescinds from explicit boundary conditions, it might be useful for dealing with Coulomb breakup problems that appear, for instance in electron-impact ionization \cite{2004McCurdy} or in breakup reactions important in astrophysics \cite{2005Alt,2012Capel} or in studying the three-body Coulomb breakup reaction of $^{11}$Li \cite{2013Kikuchi}.

In section \ref{sec.formalism} we develop the method in which the coupled Schr\"odinger equations are expanded in the  continuum bases of uncoupled channels. The first 
application of the method is done in section \ref{sec.application_PT}. It solves the problem of the exactly solvable two-channel Poeschl-Teller potential. This  works as a test case. It shows the reliability of the method and it shows the relative importance of the continuum for the deep and for the loosely bound states. In Section \ref{sec.application_deuteron}, the method is applied to the bound and scattering states of the deuteron. The last application in section \ref{sec.application_NT} is devoted to the simultaneous calculation of bound and resonant states. The outline for next applications and some remarks are given in the last section \ref{sec.conclusions}.

\section{Formalism} \label{sec.formalism}
Let us denote by $H$ the Hamiltonian which describes a  collision between two nuclei being in bound states $(a,A)$. We split $H$ into two parts: (1) the Hamiltonian $H'_\alpha$ that is left when the two initial fragments 
are far away from each other and (2) $V=\sum_{i\in a, j \in A } \, V_{ij}$ which includes the projectile($a$)-target($A$) interaction. Changing in $H'_\alpha$ to relative coordinates in each fragments and then changing to the relative coordinates between the fragments \cite{1960lane}, we end up with $H'_\alpha=H_\alpha+T$ (we have set to zero the centroid kinetic energy), where $T=- \frac{\hbar^2}{2\mu} \, \nabla^2_{r}$ is the relative kinetic energy, $\mu$ is the projectile-target 
reduced mass, and $H_\alpha= H_a + H_A$, 
where $H_a$ and $H_A$ are the intrinsic Hamiltonians of the projectile and target, respectively. Then, the total Hamiltonian reads $H=H_\alpha+T+V$. The residual interaction $V=V_d+V_{od}$ is split into a diagonal part $V_d$ and an off-diagonal one $V_{od}$ \cite{1965Tamura}. The eigenfunction $\psi_{J^\pi M}$ of $H$ is expanded into different channels using the channel basis functions $\Phi_\alpha^{J^\pi M}$ defined as

\begin{equation}
 \Phi_\alpha^{J^\pi M}(\hat{r},a,A) = [\mathcal{Y}^{j}_{l J_a}(\hat{r},a) \, \phi_{J_A}(A)]_{J^\pi M}
\end{equation}
where $\alpha=\{ (l J_a)j,J_A \}$, 
$\mathcal{Y}^{jm}_{l J_a}(\hat{r},a) = [Y_l(\hat{r}) \phi_{J_a}(a)]_{jm}$, 
$H_a \phi_{J_a M_a} = \varepsilon_a  \phi_{J_a M_a}$, 
$H_A \phi_{J_A M_A} = \varepsilon_A  \phi_{J_A M_A}$, 
$H_\alpha \phi_\alpha = \varepsilon_\alpha \phi_\alpha$, and
$\varepsilon_\alpha = \varepsilon_a + \varepsilon_A$. 

Then,

\begin{equation}\label{eq.psi}
 \psi_{J^\pi M}(\bs{r},a,A) = \sum_{\alpha'} \frac{u^{J^\pi M}_{\alpha'}(r)}{r} \, \Phi_{\alpha'}^{J^\pi M}(\hat{r},a,A)
\end{equation}

Substituting the channel expansion (\ref{eq.psi}) into the Schr\"odinger equation\\ $H\, \psi_{J^\pi M}(\bs{r},a,A) = E\, \psi_{J^\pi M}(\bs{r},a,A)$ and projecting into a certain channel $\Phi_\alpha$ we get (omitting the index ${J^\pi M}$)

\begin{equation}\label{eq.ce}
 (\varepsilon_\alpha + h_\alpha - E) \, u_\alpha(r) + 
             \sum_{\alpha' \ne \alpha} \, V_{\alpha \alpha'}(r) \, u_{\alpha'}(r) = 0
\end{equation}
where we have separated the diagonal matrix elements $V_{\alpha \alpha}$ and we have defined the single particle channel Hamiltonians,

\begin{equation}\label{eq.h}
 h_\alpha = -\frac{\hbar^2}{2 \mu} \frac{d^2}{dr^2} + 
             \frac{\hbar^2}{2 \mu} \frac{l_\alpha(l_\alpha+1)}{r^2} + V_{\alpha \alpha}
\end{equation}
with $V_{\alpha \alpha'} = \langle \Phi_\alpha | V | \Phi_{\alpha'} \rangle_{\hat{r} a A}$, where the suffix indexes mean integration over the angular coordinate $\hat{r}$ of the relative motion and the internal coordinates of the projectile $a$ and target $A$ nuclei, respectively. Notice that the structure of the Eqs. (\ref{eq.ce}) and (\ref{eq.h}) are the same to that of Eq. (25) of Ref. \cite{1965Tamura}.

Although, in principle any complete set of states will allow the computation of the interaction matrix elements, in practice, a judicious choice of the basis states will minimize the number of matrix elements to be calculated and reduce the computation time needed.  Here we use the diagonal part $V_d$ of the residual interaction $V=V_d + V_{od}$ to generate the basis. Notice that the basis does not correspond to the one generated without residual interaction $V=0$.

In the next step, we expand the wave functions $u_\alpha(r)$ in each channel in the basis generated by its own channel Hamiltonian $h_\alpha$

\begin{equation}
 h_\alpha \,u^{(0)}_{\alpha, n}(r) = \varepsilon^{(0)}_{\alpha, n} \, u^{(0)}_{\alpha, n}(r)
\end{equation}
\begin{equation}\label{eq.uexp}
 u_{\alpha'}(r) = \sum_{n'} \, c_{\alpha', n'} \, u^{(0)}_{\alpha', n'}(r)
\end{equation}
where the summation  includes integration over the continuum part of the spectrum of $h_\alpha$.

Replacing the expansion of $u_{\alpha}(r)$ (Eq. (\ref{eq.uexp})) in Eq. (\ref{eq.ce}) and projecting over $u^{(0)}_{\alpha, n}(r)$ we get

\begin{eqnarray}\label{eq.ce2}
 \sum_{\alpha'=1}^N \sum_{n'=1}^{M_{\alpha'}} \, \left[
                        (\varepsilon_\alpha + \varepsilon^{(0)}_{\alpha,n} - E) \, 
                                           \delta_{\alpha \alpha'} \, \delta_{nn'} + \right. 
               \left.         (1-\delta_{\alpha \alpha'}) \, V_{\alpha\, n, \alpha' \, n'}
                      \right] \, c_{\alpha',n'}  = 0
\end{eqnarray}
where $N$ denotes the number of channels and $M_\alpha$ is the number of single particle basis states for the channel $\alpha$. 

The coupled equations problem in Eq. (\ref{eq.ce2}) can be transformed to an eigenvalue problem with a sparse symmetric matrix of dimension $M=M_1+...+M_N$ by defining the index $i=\{ \alpha,n \}$ with the following order $i=\{ (\alpha_1,1)$, $(\alpha_1,2)$, $\dots$, $(\alpha_1,M_1)$, $(\alpha_2,1)$, $\dots$, $(\alpha_2,M_2)$, ...,$(\alpha_N,1)$, $\dots$, $(\alpha_N,M_N) \}$. The matrix is diagonal in each channel block $\alpha$ of dimension $M_\alpha$. The diagonal elements in each channel block $\alpha$ are given by $\varepsilon_\alpha + \varepsilon^{(0)}_{\alpha,n}-E$, with $n=\{1,2,...,M_\alpha \}$. The matrix elements between different channels contain only the interaction $V_{i i'}$ given by

\begin{eqnarray}
 V_{\alpha\, n, \alpha' \, n'} &=& \int \, dr \, u^{(0)}_{\alpha,n}(r) \, V_{\alpha \alpha'}(r) \,
                                                 u^{(0)}_{\alpha',n'}(r)  \nonumber 
\end{eqnarray}

Using the basis generated by the diagonal part of the channel interaction 
one can save the calculation of $M_{saved}=\sum_{\alpha=1}^{N} \, \frac{M_\alpha(M_\alpha+1)}{2}$ interaction matrix elements. The number of these matrix elements increases rapidly as the number of open channels $N$ and the  dimension  of the 
basis $M_\alpha$ increases.

There are two advantages of using a basis expansion method instead of using the asymptotic boundary conditions.  The matrix diagonalization does not diverge even if the coupling terms are large. This might happen in the direct numerical integration \cite{1988Thompson} of the coupled 
equations. The matrix diagonalization does not face any instability of the numerical integration of the coupled equations. The disadvantage of using a basis expansion is that one has to deal with the completeness problem of the basis. A difficulty of using basis expansion is that one 
needs an efficient and accurate method to solve the single particle Schr\"odinger 
equation, that is,  
to find real and complex poles as well as the real and complex energy scattering states. The real and complex energy scattering states were calculated 
by using a piecewise perturbation method \cite{1995Ixaru}. The code implements the so called Ixaru's method \cite{1984Ixaru}. The real and complex energy poles were also calculated by using a modified version of the program \cite{1995Ixaru}. This version has a higher precision than the GAMOW code \cite{1982Vertse}, which however is more 
flexible.

\section{Application to the Poeschl-Teller potential: bound state calculation using bases composed of bound states and real energy continuum} \label{sec.application_PT}
In this section we compare the exact solution of the two-channel Poeschl-Teller potential with the numerical solution using the same eigenchannel 
bases for both channels. The bases are composed of bound and real energy scattering states.

Let us consider the Schr\"odinger equations with two-channels and with $\hbar=2\mu=1$, $l_1=l_2=0$, $\varepsilon_1=\varepsilon_2=0$ and $V_{\alpha \alpha'}(r)$ given by Ixaru \cite{2008Ixaru}

\begin{equation}\label{eq.vi}
 V_{\alpha \alpha'}(r)=\left(
            \begin{array}{cc}
              V_d(r) & V_{od}(r) \\
              V_{od}(r) & V_d(r)
            \end{array}
      \right)
\end{equation}

For this special interaction the coupled equations (\ref{eq.ce}) can be collected into two uncoupled equations,

\begin{eqnarray} \label{eq.pt}
 \left[ h^+(r) - E^+ \right] u^+(r) &=& 0 \label{eq.pt1} \\
 \left[ h^-(r) - E^- \right] u^-(r) &=& 0 \label{eq.pt2} 
\end{eqnarray}
with $h^\pm(r) = -\frac{d^2}{dr^2} + V^\pm(r)$ and $V^\pm(r)=V_d(r) \pm V_{\rm{od}}(r)$. Then, one may choose $V_d$ and $V_{\rm{od}}$ such that $V^\pm$ have exact solutions. In this way we find the eigenvalues $E^+$ and $E^-$ of $h^+$ and $h^-$ which will be also eigenvalues of the original coupled-equation, i.e. $E=\{ E_1^+,E_2^+,...,E_1^-,E_2^-,... \}$.

Taking \cite{2008Ixaru},

\begin{eqnarray} 
 V_d(r) &=& V_{PT}(r;-45,1) + V_{PT}(r;-\frac{39}{2},\frac{1}{2}) \label{eq.via} \\
 V_{od}(r) &=& V_{PT}(r;-45,1) - V_{PT}(r;-\frac{39}{2},\frac{1}{2}) \label{eq.vib}
\end{eqnarray}
one gets $V^+(r)=V_{PT}(r;-90,1)$ and $V^-(r)=V_{PT}(r;-39,\frac{1}{2})$ where\\ $V^\pm_{PT}(r;V^\pm_0,\alpha^\pm)=V^\pm_0\,cosh^{-2}(\alpha^\pm r)$ is the Poeschl-Teller potential with eigenenergies $E^\pm_n=-4\, (\alpha^\pm)^2\, (n-t^\pm)^2$, $n=0,1,...,n_{\rm{max}}$; with $n_{\rm{max}}$ the largest integer smaller than $t^\pm$, $t^\pm=0.25[-3+\sqrt{1-4 V^\pm_0/(\alpha^\pm)^2}]$.

\subsection{Basis expansion}
\begin{figure}[htb]
\vspace{5mm}
\begin{center}
 \includegraphics[width=0.55\textwidth]{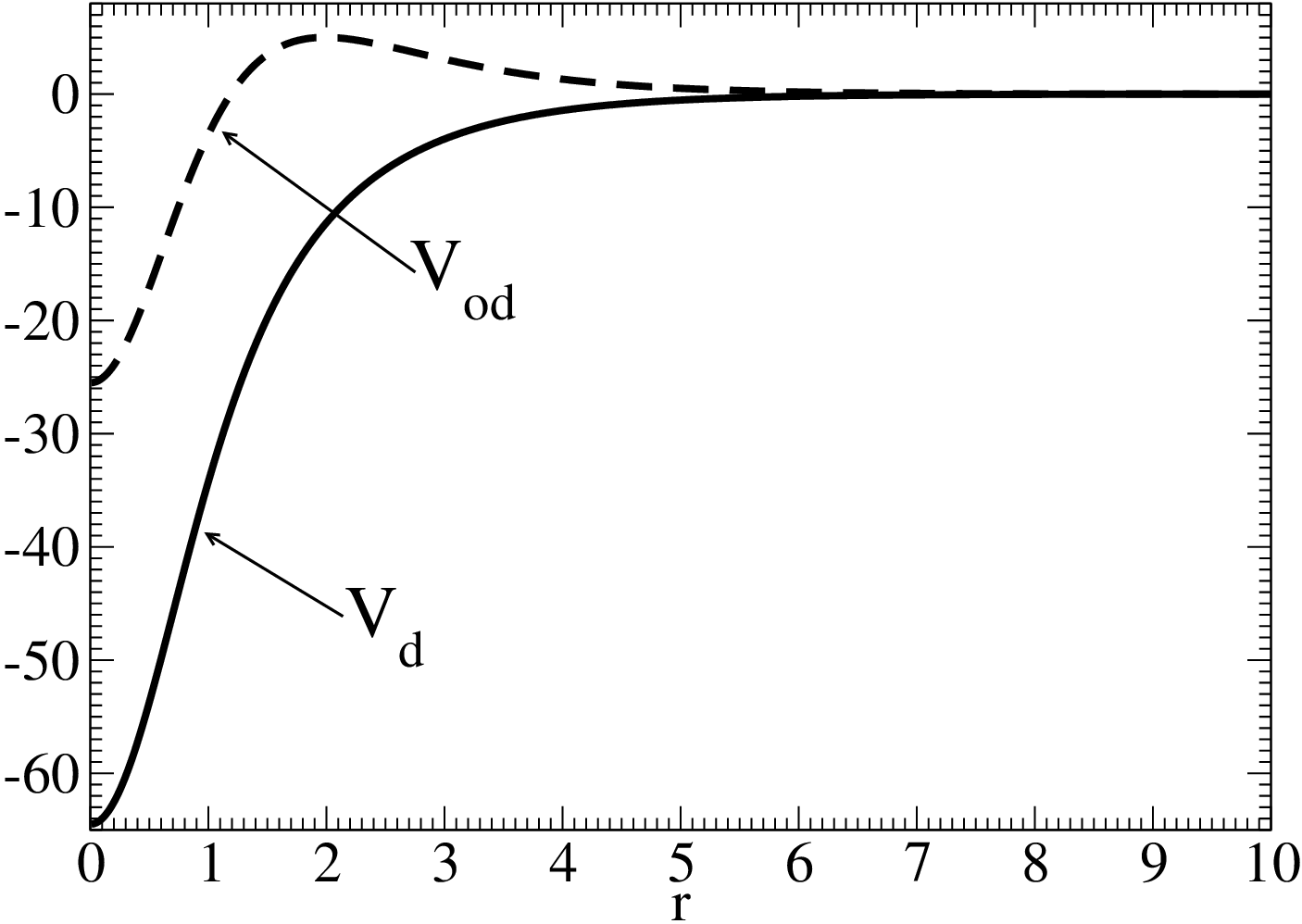}
 \caption{\label{fig.pt} Diagonal $V_d(r)$ and off-diagonal $V_{\rm{od}}(r)$ parts of the two-channel Poeschl-Teller potential.}
\end{center}
\vspace{5mm}
\end{figure}

The channel wave functions $u_1(r)$ and $u_2(r)$ in Eq. (\ref{eq.ce}) are expanded in the same basis, since $h_1=h_2=-\frac{d^2}{dr^2} + V_d(r)$. The potential $V_d(r)$ Eq. (\ref{eq.via}), is shown if Fig. \ref{fig.pt}. The basis is formed by the five bound states $u^{(0)}_i(r)$, with $i=1,\dots,5$ and $N_c$ real energy scattering states $u^{(0)}(r,\varepsilon_j)$ with $j=1,\dots,N_c$, of $h_1$.
The energies of the bound states are: 
$\varepsilon^{(0)}_{1}=-45.5475$, 
$\varepsilon^{(0)}_{2}=-26.1325$, 
$\varepsilon^{(0)}_{3}=-13.0841$, 
$\varepsilon^{(0)}_{4}=-5.24900$, 
$\varepsilon^{(0)}_{5}=-1.25622$, while the continuum is discretized using Gauss-Legendre partition $\varepsilon_i \in (0,\varepsilon_{\rm{max}})$ with weights $\omega_i$.

The matrix elements $V_{\alpha n,\alpha' n'}$ (with $\alpha,\alpha'=1,2$ and $n,n'=1,\dots,5+N_c$ ) were calculated using the potential $V_{\rm{od}}(r)$ Eq. (\ref{eq.vib}) shown in fig. \ref{fig.pt}. The integration was performed by using Gauss-Legendre quadrature with $40$ mesh points between $r=(0,10)$.

The convergence of the solutions was studied as a function of the  
cutoff  energy $\varepsilon_{\rm{max}}$ and the number of mesh points $N_c$. Table \ref{table.pt} shows the convergence of the energies for $\varepsilon_{\rm{max}}=70$ as the function 
of the number of continuum states $N_C$ . We can see a fast convergence for all states except the state being closest to the threshold. It is 
worthwhile to mention that all ten perturbed bound states were found in a single diagonalization by using  bases with only five (unperturbed) bound states.

\begin{table}[ht]
\begin{center}
\small{
 \caption{\label{table.pt} Two-channels Poeschl-Teller energies obtained in diagonalization as function of the number $N_c$ of continuum states in the bases. The cutoff energy is $\varepsilon_{\rm{max}}=70$. $E_{\rm{exact}}$ refers to the exact energies of Eqs. (\ref{eq.pt1}) and (\ref{eq.pt2}). The last column shows the relative error $e_{\rm{rel} }$ in $\%$ for $N_c=70$.}
 \begin{tabular}{c|ccccc|c|c}
 \hline
            &          &          &   $N_c$  &          &          &                  & \\ 
  State $n$ &  0       &   10     &   30     &   50     &  70      & $E_{\rm{exact}}$ & $e_{\rm{rel}}\%$ \\ 
 \hline
  $E_1$     & -63.962  & -63.999  & -63.999  & -63.999  & -63.999  & -64.000          & 0.002\\
  $E_2$     & -35.425  & -35.982  & -35.988  & -35.990  & -35.990  & -36.000          & 0.028 \\
  $E_3$     & -30.249  & -30.250  & -30.250  & -30.250  & -30.250  & -30.250          & 0.000 \\
  $E_4$     & -20.240  & -20.250  & -20.250  & -20.250  & -20.250  & -20.250          & 0.000 \\
  $E_5$     & -14.289  & -15.944  & -15.967  & -15.970  & -15.971  & -16.000          & 0.181 \\
  $E_6$     & -12.160  & -12.240  & -12.242  & -12.243  & -12.244  & -12.250          & 0.049 \\
  $E_7$     & -5.0767  & -6.1836  & -6.1951  & -6.2025  & -6.2043  & -6.2500          & 0.731 \\
  $E_8$     & -2.6000  & -3.9565  & -3.9786  & -3.9801  & -3.9804  & -4.0000          & 0.490 \\
  $E_9$     &  0.55301 & -2.2636  & -2.2478  & -2.2482  & -2.2482  & -2.2500          & 0.080 \\
  $E_{10}$  &  0.91049 &  0.18570 & -0.11812 & -0.14446 & -0.15007 & -0.25000         & 39.972 \\
  \hline
\end{tabular} 
}
\end{center}
\end{table}

\section{Application to the proton-neutron system: bound state and eigenphase shifts calculations using real energy continuum bases}\label{sec.application_deuteron}
In this section we solve the coupled-equation for the deuteron, and calculate the eigenphases $\delta_S$ and $\delta_D$ using the soft core potential of Ref. \cite{1968Reid}.  Only a single adjustable parameter, the cutoff  energy is used here. 

\begin{figure}[htb]
\vspace{5mm}
\begin{center}
 \includegraphics[width=0.55\textwidth]{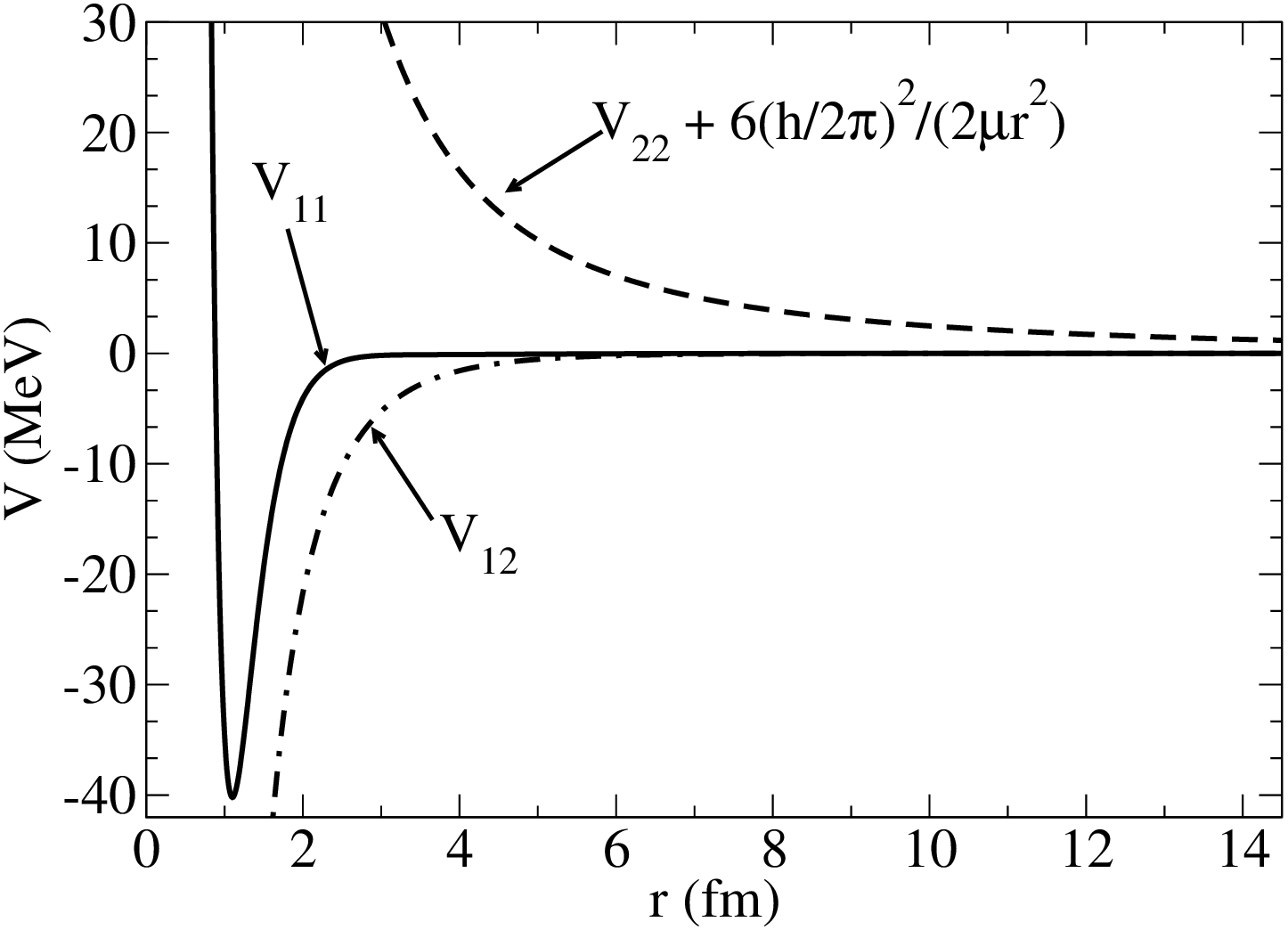}
 \caption{\label{fig.pot} First channel diagonal potential $V_{11}$, Eq. (\ref{eq.reid1}) (continuum line); second channel diagonal potential $V_{22}$, Eq. (\ref{eq.reid2}) plus centrifugal potential (dashed line); and off-diagonal potential $V_{12}$, Eq. (\ref{eq.reid3}) (dash-point line) for the triple-even $np$ state.}
\end{center}
\vspace{5mm}
\end{figure}

For this system the quantities which appear in the coupled equations (\ref{eq.ce}) are: $l_1=0$, $l_2=2$, $\varepsilon_1=\varepsilon_2=0$, $\frac{1}{\mu}=\frac{1}{m_p}+\frac{1}{m_n}$ (with $m_n$ and $m_p$ the neutron and proton mass, respectively), and $2 \mu/\hbar^2=0.0241138\,(MeV.fm^2)^{-1}$. The potentials (see fig. \ref{fig.pot}) are given by the following expressions,

\hspace{6mm}
\begin{eqnarray} 
 V_{11} &=& V_C(r) \label{eq.reid1} \\
 V_{22} &=& V_C(r) - 2\, V_T(r) - 3\, V_{LS}(r) \label{eq.reid2}\\
 V_{12} &=& V_{21} = 2\sqrt{2}\, V_T(r) \label{eq.reid3}
\end{eqnarray}
with

{\footnotesize
\begin{eqnarray}
 V_C(r) &=& -10.463\, \frac{e^{-x}}{x} + 105.468\, \frac{e^{-2x}}{x} - 3187.8\, \frac{e^{-4x}}{x} + 9924.3\, \frac{e^{-6x}}{x} \label{eq.reid4} \\
 V_T(r) &=& -10.463\, \left[ \left( 1 + \frac{3}{x} + \frac{3}{x^2} \right) \, \frac{e^{-x}}{x}
                        \left( \frac{12}{x} + \frac{3}{x^2} \right) \, \frac{e^{-4x}}{x}
                 \right] \nonumber \\
           && + 351.77\, \frac{e^{-4x}}{x} + 1673.5\, \frac{e^{-6x}}{x} \label{eq.reid5} \\
 V_{LS}(r) &=& 708.91\, \frac{e^{-4x}}{x} - 2713.1\, \frac{e^{-6x}}{x}~. \label{eq.reid6} 
\end{eqnarray}
}
The radial coordinate $r$ is given in fm while the interactions are given in MeV units and $x=(0.7fm^{-1})\times r$ is dimensionless.

For ground state of the deuteron, the channel wave functions $u_1(r) \equiv u_S(r)$ and $u_2(r) \equiv u_D(r)$ correspond to the $^3S_1$ and $^3D_1$ components of the wave function. While for the $n-p$ scattering states $u_1(r)$ and $u_2(r)$ are standing-waves \cite{1983Barret} which asymptotically behave like $u_{1,\alpha}(r) \sim sin(kr + \delta_\alpha)$ and $u_{2,\alpha}(r) \sim sin(kr - \pi + \delta_\alpha)$ with $\alpha=S,D$ and $\delta_S$ and $\delta_D$ the eigenphase shifts \cite{1952Blatt,1973Barret}. 

\subsection{Basis expansion}
Since none of the diagonal potentials hold any bound state (the potential $V_{11}$ has an anti-bound state at the energy $-5.671$ MeV), the two bases are formed from discretized continuum states $\sqrt{\omega_i} \, u^{(0)}_{\alpha}(r,\varepsilon_i)$ only. Here $\omega_i$ are the weights of the Gauss-Legendre mesh points at the energies $\varepsilon_i \in (0,\varepsilon_{\rm{max}})$. We took the same continuum basis for both channels, i.e. both channels were expanded using the same number of mesh points up to the same cutoff   energy $\varepsilon_{\rm{max}}$. 

The interaction matrix elements were calculated using Gauss-Legendre quadrature with $r \in (0,r_{\rm{max}})$. It was checked that the bound state solution was stable when we varied the cutoff radius between $16$ fm to $24$ fm. For the calculation we took $20$ fm for the  cutoff radius and $100$ for the number of mesh points.

\begin{figure}[htb]
\vspace{5mm}
\begin{center}
 \includegraphics[width=0.55\textwidth]{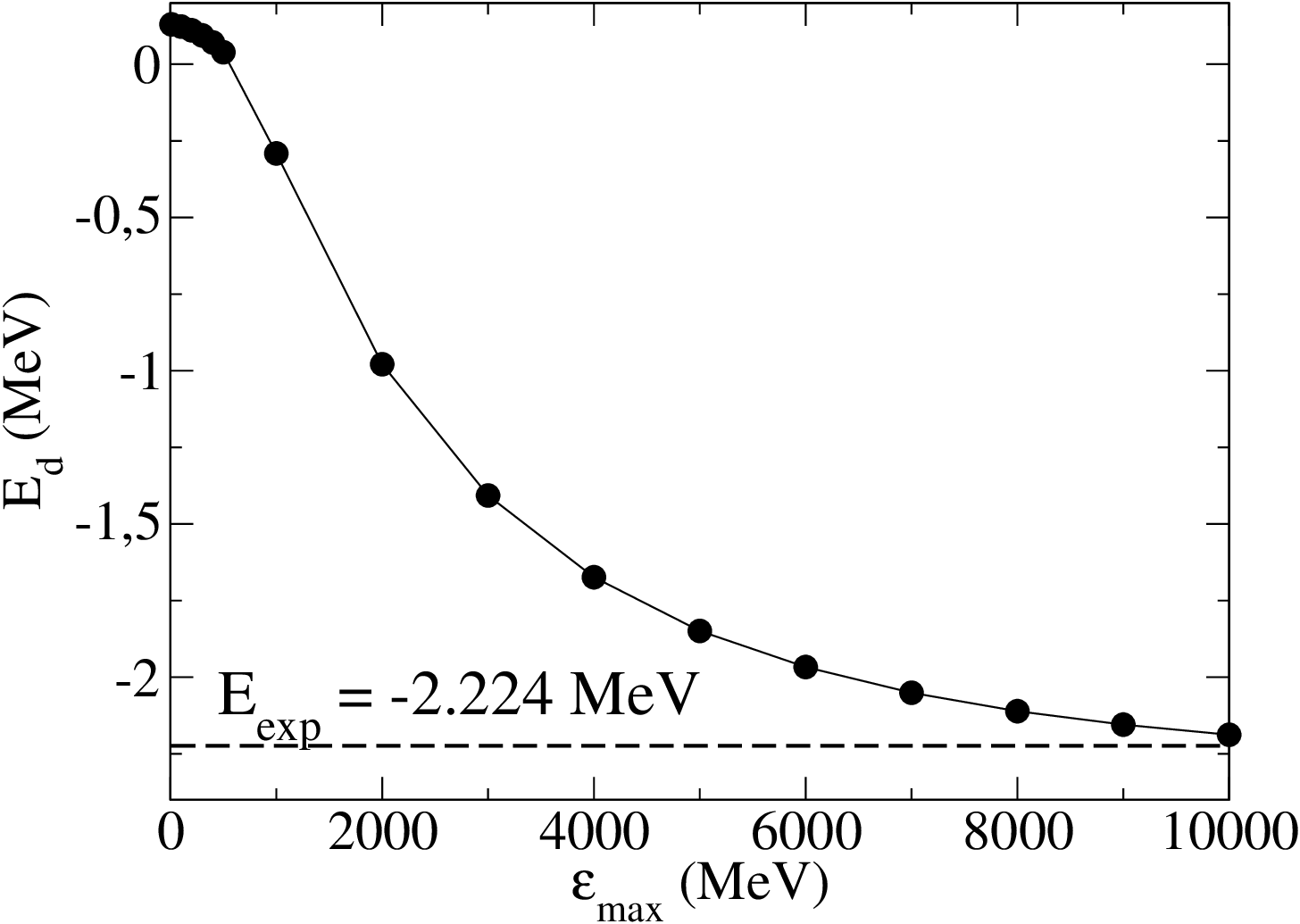} 
 \caption{\label{fig.enp} Deuteron ground state energy $E_{\rm{d}}$ as a function of the  cutoff energy  $\varepsilon_{\rm{max}}$. The big dots represent the results of numerical calculations, while the thin line is just to guide the sight. The dashed horizontal line represents the experimental ground state energy.}
\end{center}
\vspace{5mm}
\end{figure}

The ground state energy $E_{\rm{d}}$ and wave function of the deuteron were calculated as a function of the cutoff  energy $\varepsilon_{\rm{max}}$. The real 
energy basis states were defined by the following vertices (in MeV): $(0,10)$, $(10,50)$, $(50,100)$, $(100,250)$, $(250,500)$, $(500,750)$, $(750,1000)$, $(1000,2000)$,\dots $(9000,10000)$, where $\dots$ means that 
an interval of $1000$ MeV have been used. Six mesh points for each interval up to $1000$ MeV and ten mesh points from there on, have been taken. While the same energy 
partition was used for both channels the scattering functions were not the same since $h_1$ and $h_2$ were different. Figure \ref{fig.enp} shows that the deuteron ground state energy converges very slowly to the experimental value $E_{\rm{exp}}=-2.224$ MeV as the cutoff energy increases. The final value $E_{\rm{d}}=-2.210$ MeV gives  93.7 \% and 6.3 \%  for the $^3S_1$ and $^3D_1$ partial wave amplitudes, respectively. Both, the energy and the wave function content fit well to the experimental values.

\begin{figure}[htb]
\vspace{5mm}
\begin{center}
 \includegraphics[width=0.45\textwidth]{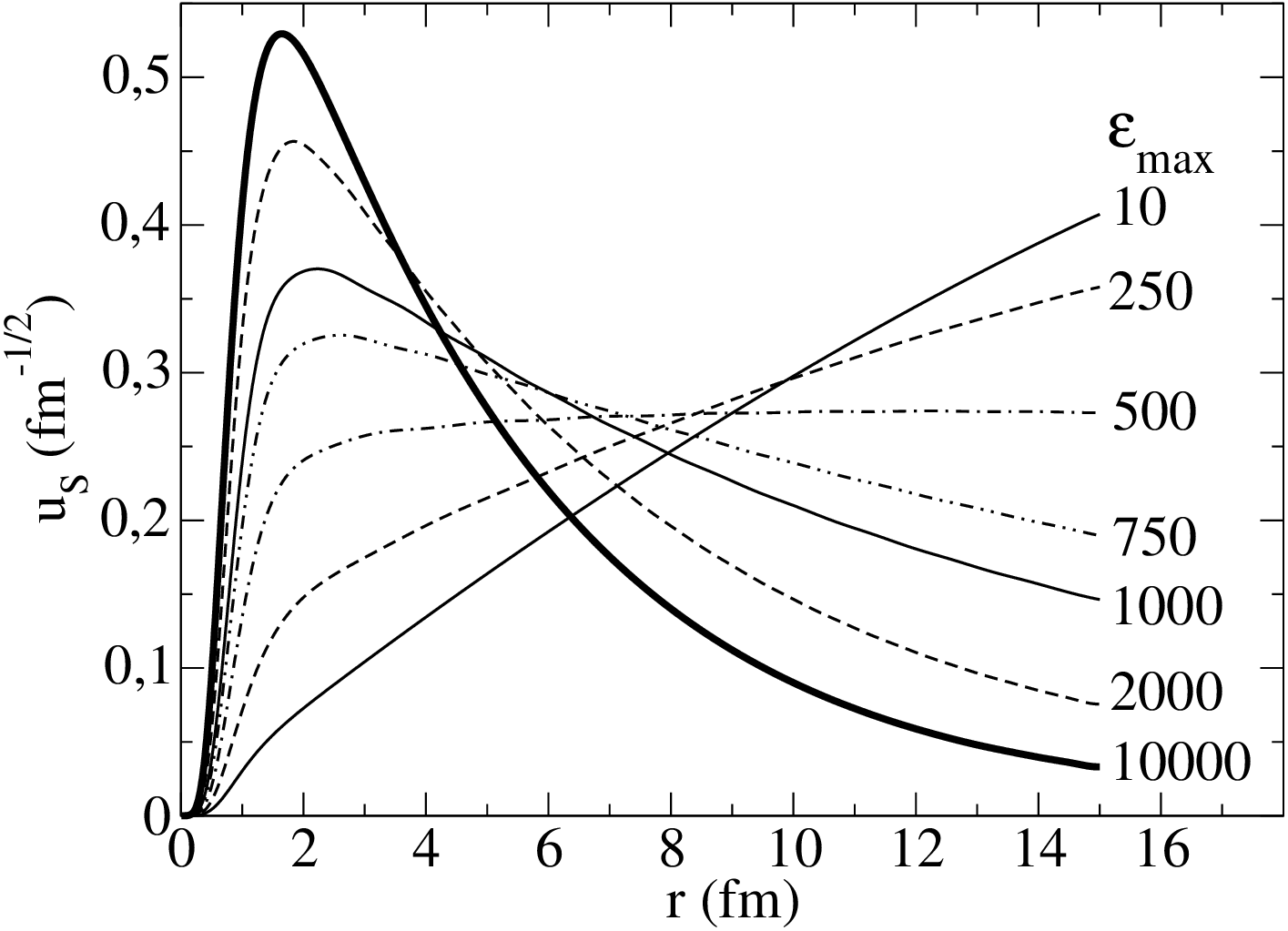}
 \hspace{5mm}
 \includegraphics[width=0.45\textwidth]{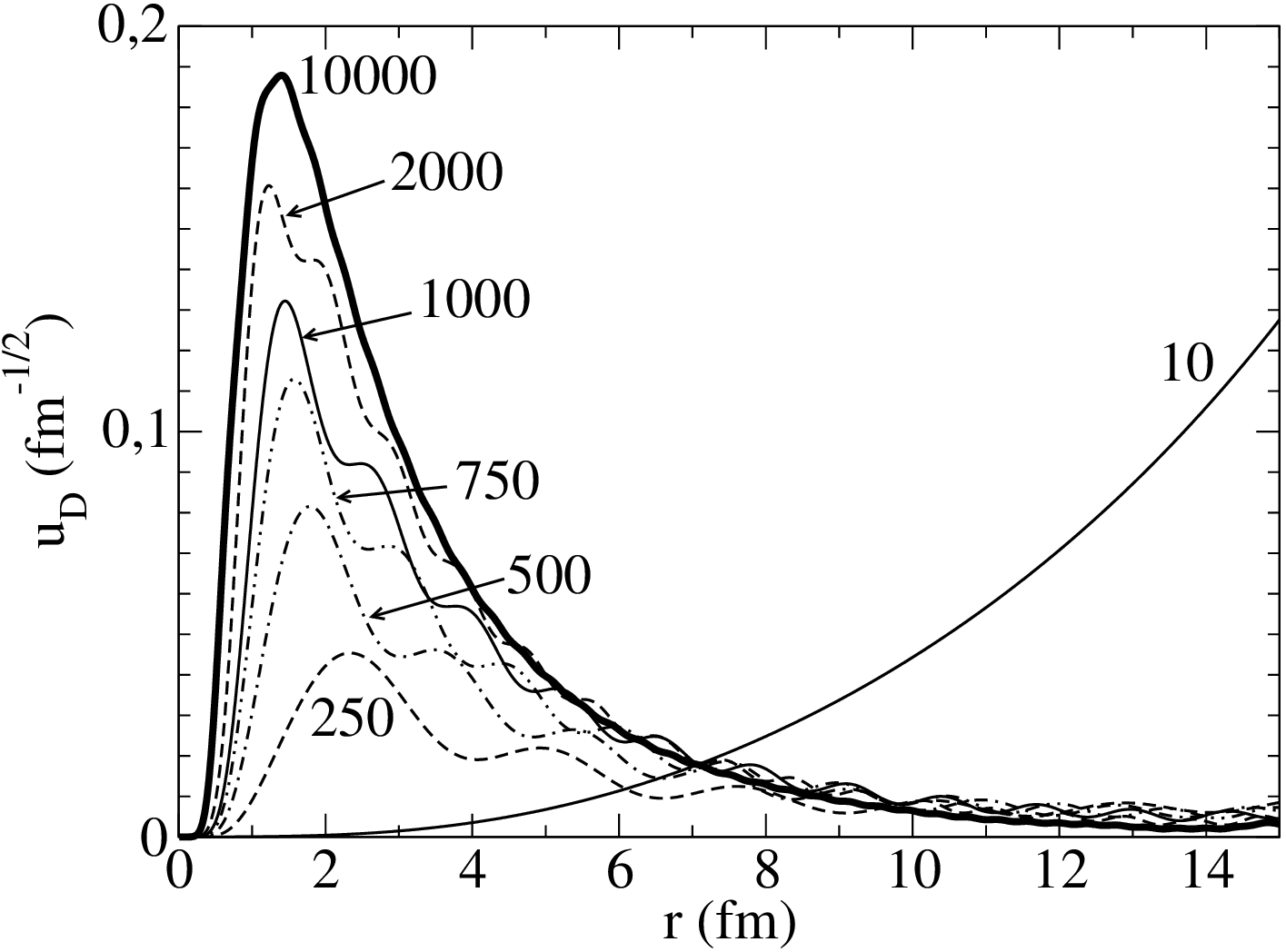}
 \caption{\label{fig.wfnp} Deuteron ground state wave function components $u_S$ (left) and $u_D$ (right) parameterized in the cutoff energy $\varepsilon_{\rm{max}}$. The labels correspond to the value of $\varepsilon_{\rm{max}}$. The oscillations due to the incomplete basis are sleeked as the cutoff energy increases.}
\end{center}
\vspace{5mm}
\end{figure}

The ground state wave function was built from the eigenvector of 
$E_{\rm{d}}$,  Eq. (\ref{eq.uexp}). Fig. \ref{fig.wfnp} shows the convergence of the channel wave functions as the cutoff energy  increases. The $D$ component of the deuteron wave function shows oscillations due to the oscillations in the high energy basis states. The magnitude of the oscillations decreases by increasing the mesh points energies or by increasing the cutoff energy. In short, the oscillations is due to the incomplete representation used in the expansion of Eq. (\ref{eq.uexp}). The oscillations in the $S$ component are much smaller and they are not visible at the scale used in the figure. The difference in the magnitudes of the oscillations between the two components of the wave function could be attributed to the large differences between the values of the diagonal potentials in the tail region (see fig. \ref{fig.pot}).

\begin{figure}[htb]
\vspace{5mm}
\begin{center}
 \includegraphics[width=0.55\textwidth]{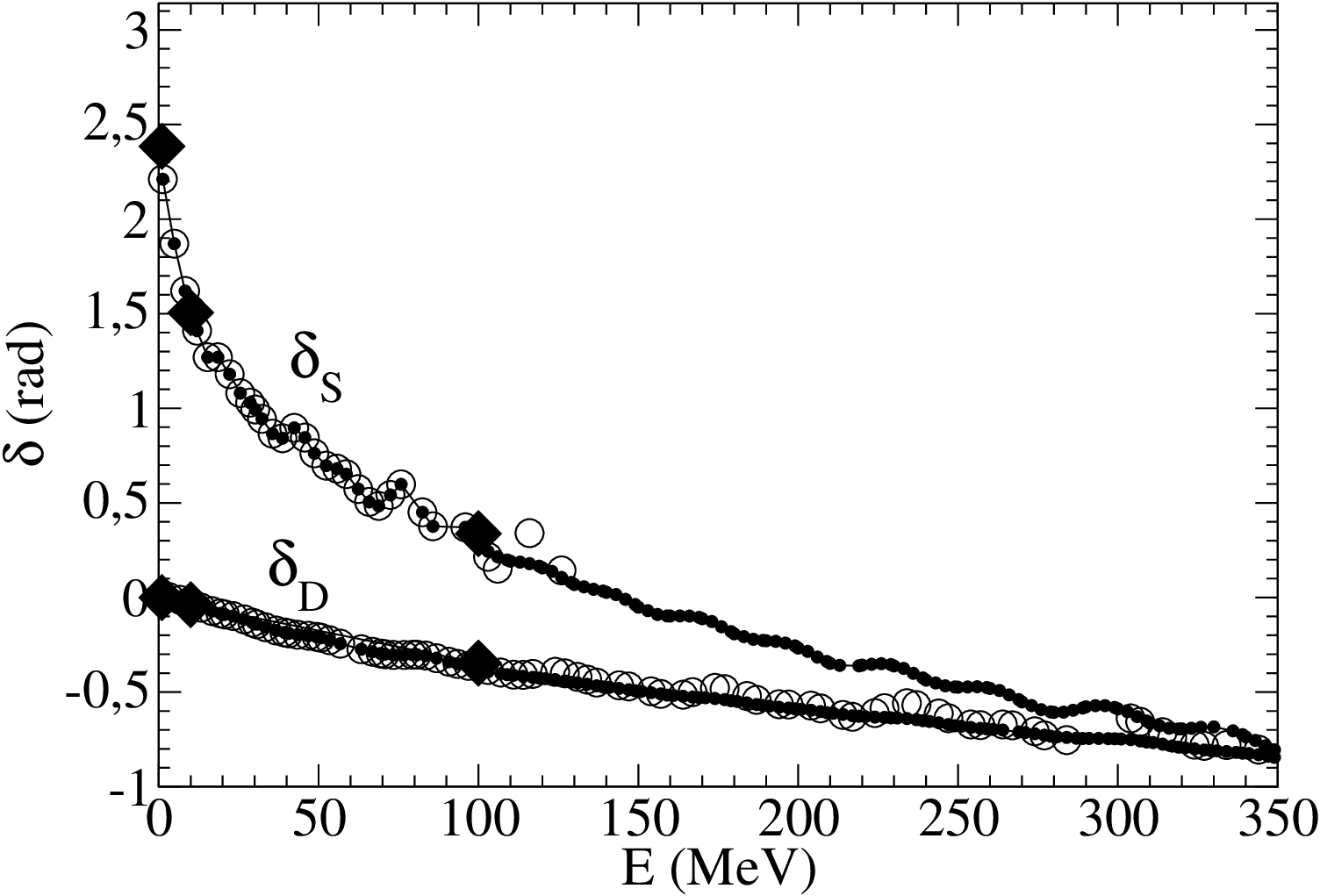}
 \caption{\label{fig.ps} $\delta_S$ and $\delta_D$ eigenphases in the proton-neutron scattering using the soft Reid potential. The big open circle are the calculation fitting all three parameters of the asymptotic $A[cos(\delta)*F_0(kr) + sin(\delta)*G_0(kr)] + B[cos(\delta)*F_2(kr) + sin(\delta)*G_2(kr)]$, while the small filled dots joined by a thin line are the eigenphases found fitting the eigenphase shift and the main component of the scattering wave function. The large diamond filled symbol are the results from Ref. \cite{1998Hesse} using $R$-matrix formalism.}
\end{center}
\vspace{5mm}
\end{figure}

From the diagonalization we obtained, beside the ground state, the scattering states. As for the bound ground state, we get the corresponding eigenvectors for each positive eigenvalue. Then, Eq. (\ref{eq.uexp}) gives the scattering states expanded in the continuum basis. These scattering solutions can be used to calculate the eigenphase shifts \cite{1952Blatt,1973Barret} $\delta_S$ and $\delta_D$. From each scattering state we built the sum of the two channels wave functions and fit it to the function $A[cos(\delta)*F_0(kr) + sin(\delta)*G_0(kr)] + B[cos(\delta)*F_2(kr) + sin(\delta)*G_2(kr)]$, for $r > 9$ fm. The functions $F_l(kr)$ and  $G_l(kr)$ are the regular and irregular Coulomb functions, respectively. They were calculated using the program \cite{1985Thompson}. The eigenphases were calculated using the Levenberg-Marquad code from Numerical Recipes \cite{1996nr}. Using the same basis that for the deuteron we calculated the eigenphase shifts $\delta_S$ and $\delta_D$ and compared them with the results obtained by doubling the mesh points for energy larger than $1000$ MeV; it was found that the values of $\delta_S$ changed around 1\% while the values of $\delta_D$ changed in the third figure. Then, we doubled and tripled the mesh for the energies below  $1000$ MeV. There, it was found that the eigenphase $\delta_D$ has a smooth behavior, while the eigenphase $\delta_S$ shows oscillations around and above $100$ MeV. In order to have a more uniform distribution, we made intervals of $10$ MeV from zero up to $350$ MeV and took the same number of point $n_c$ in each interval. We calculated the eigenphases for increasing $n_c$. For $n_c=1,2,3$, we found the same qualitative behavior than for the deuteron basis, while for $n_c=4$ almost no $\delta_S$ was found for energies larger that $120$ MeV within the required error. Instead $\delta_D$ was fitted all the range except between $290$ MeV and $310$ MeV. Using these last basis the fit was done with the restriction that for energies larger than $100$ MeV only the amplitude of the most important channel together with the eigenphase shift was fitted. Figure \ref{fig.ps} shows the results of these two calculations. We can notice that $\delta_D$ values resulted by both calculations are very similar, while the two parameters fit calculation smoothly connects to the calculation of $\delta_S$ using the three parameters fit.

\section{Application to the Noro-Taylor potential: resonant state calculations using complex energy basis states}\label{sec.application_NT}
The Noro-Taylor potential \cite{1980Noro} is a two-channel model with a strong repulsion in the second channel and with a strong coupling. This system has a narrow resonance at the energy $E_r=4.7682$ ($\Gamma=0.001420$). The parameters (in atomic units) which appear in the coupled equations (\ref{eq.ce}) are: $l_1=l_2=0$, $\varepsilon_1=0$, $\varepsilon_2=0.1$, $\mu=\hbar=1$. The potentials are given by the following expressions,

\begin{eqnarray} 
 V_{11} &=& -\,r^2 e^{-r} \label{eq.nt1} \\
 V_{22} &=&  7.5\, r^2 e^{-r} \label{eq.nt2} \\
 V_{12} &=& -7.5\, r^2 e^{-r} \label{eq.nt3} 
\end{eqnarray}

\subsection{Basis expansion}
The diagonal potential $V_{11}$ has two bound states at energies 
$\varepsilon^{(0)}_{1,1}=-0.296188$ and 
$\varepsilon^{(0)}_{1,2}=-0.0106981$, while the potential $V_{22}$ has a resonance at the energy 
$\varepsilon^{(0)}_{2,1}=(3.42639,-0.0127745)$. The matrix elements $V_{1 n,2 n'}$ were calculated using  Gauss-Legendre quadrature with $100$  mesh points for the radial coordinate from $0$ to $35$.

Since the potential $V_{22}$ is repulsive the channel wave function $u_2$ is expanded only by continuum basis states. We considered two different  bases: 
(1) only real positive energy states are included, which we call real energy representation, and 
(2) complex energy states are also included, which we call complex energy (Berggren) representation. Using the real representation only bound states can be found while using a properly chosen complex energy representation we can get also 
resonant states.

The first channel wave function $u_1$ is expanded 
in terms of the wave functions of  the two bound states $\varepsilon^{(0)}_{1,1}$ and $\varepsilon^{(0)}_{1,2}$ plus a set of discretized continuum states along the real axis up to a cutoff energy $\varepsilon_{1,\rm{max}}$. The second channel wave function $u_2$ can either be expanded by using a set of discretized real energy scattering states up to an cutoff energy $\varepsilon_{2,\rm{max}}$ (real energy representation) or alternatively by the resonant state $\varepsilon^{(0)}_{2,1}$ plus a set of discretized complex energy scattering states taken along a contour in the complex energy plane. 

The resonance in the second channel affects the selection of the mesh points since its presence requires a denser mesh of the scattering states in the vicinity of the resonant energy. The second column of table \ref{table.cnt} shows the contour used in the real energy representation (RR), while the fourth column shows the contour for the complex energy representation (CR). Notice that the real representation requires a larger $\varepsilon_{2,\rm{max}}$.

\begin{table}[htb]
\begin{center}
 \caption{\label{table.cnt} Vertexes of the contours and the number of mesh-points $N$ for the real (RR) and complex (CR) representations for the Noro-Taylor potential.}
\begin{tabular}{c|cc|cc} 
         & RR       &    & CR &\\
 \hline
 Channel & Vertex   & N  & Vertex     & N  \\
 \hline
  1      & (50.,0.) & 50 & (50.,0.)   & 50 \\
 \hline
  2      & (3.4,0.) & 50 & (3.4,-0.1) & 20 \\
  2      & (3.5,0.) & 50 & (4.7,-0.1) & 20  \\
  2      & (250,0.) & 50 & (6.8,0)    & 20\\
  2      &          &    & (50,0)     & 50
\end{tabular} 
\end{center}
\end{table}

Table \ref{table.ent} shows the perturbed energies calculated using either the real or the complex energy representations. For comparison we give the results obtained in Ref. \cite{2006Rakityansky} using the Jost function method combined with complex energy rotation, this corresponds to the last column 
(under the title $E_{\rm{exact}}$). It is found that only the complex energy representation (Berggren basis \cite{1968Berggren}) is able to reproduce simultaneously bound and unbound perturbed states in this coupled channels system. 

\begin{table}[htb]
\begin{center}
 \caption{\label{table.ent} First five poles of the Noro-Taylor potential calculated from the diagonalization in real ($E_{RR}$) and complex ($E_{CR}$) representations compared to $E_{\rm{exact}}$ given in Ref. \cite{2006Rakityansky}.}
\begin{tabular}{c|ccc} 
 $n$ & $E_{RR}$   & $E_{CR}$          & $E_{\rm{exact}}$   \\
 \hline
  1  & -2.321     & -2.316            & -2.314  \\
  2  & -1.327     & -1.312            & -1.310  \\
  3  & -0.5554    & -0.5396           & -0.5374  \\
  4  & -0.07627   & -0.06496          & -0.06526  \\
  5  &            & (4.769,-0.00075)  & (4.768,-0.00071) 
\end{tabular} 
\end{center}
\end{table}

\section{Conclusions} \label{sec.conclusions}
Loosely and deeply bound states, resonant states, and eigenphase shifts have been calculated in  
an optimized continuum basis. 

Each channel defines its own basis (eigenchannel basis) through the diagonal parts of the channel potentials, in this way the number of matrix elements to be calculated is reduced considerably.

The matrix diagonalization gives the poles and scattering solutions simultaneously, i.e. the basis is energy-independent. 

Since the solutions of the coupled-channel equations do not rely on boundary conditions, the method could be convenient for studying systems where the boundary conditions cannot be treated easily.

In summary, we presented a method for describing nuclear reactions involving weakly bound or unbound system. The next step is to apply the continuum eigenchannel basis expansion for studying deuteron elastic breakup process.

The author thanks Prof. T. Vertse for valuable discussions. This work has been partially supported by the National Council of Research PIP-77 (CONICET, Argentina).


\end{document}